\documentclass[preprintnumbers,article,amsmath,amssymb,floatfix,10pt,prd,onecolumn,
superscriptaddress,nofootinbib]{revtex4}
\usepackage[colorlinks=true, pdfstartview=FitV, linkcolor=blue, citecolor=red, urlcolor=magenta]{hyperref}
\usepackage{bbm}
\usepackage{amsfonts}
\usepackage{mathrsfs}
\usepackage{latexsym}
\usepackage{epsfig}
\usepackage{epstopdf}
\usepackage{epstopdf}
\usepackage{graphicx}
\usepackage{amssymb}
\usepackage{amsmath}
\usepackage{dcolumn}
\usepackage{bm}
\usepackage{float}
\usepackage{color}
\usepackage{comment}
\usepackage{xcolor}
\begin{document}
\title{\bf Imprints of dark energy models on structural properties of charged gravastars in extended teleparallel gravity}

\author{G. Mustafa}
\email{gmustafa3828@gmail.com} \affiliation{Department of
Physics, Zhejiang Normal University, Jinhua 321004, People's
Republic of China}

\author{Faisal Javed}
\email{faisaljaved.math@gmail.com}\affiliation{Department of
Physics, Zhejiang Normal University, Jinhua 321004, People's
Republic of China}

\author{Arfa Waseem}
\email{arfa.waseem@gcwus.edu.pk} \affiliation{Department of
Mathematics, Government College Women University, Sialkot, Pakistan}

\author{S.K. Maurya}
\email{sunil@unizwa.edu.om}\affiliation{Department of Mathematical and Physical Sciences, College of Arts and Sciences, University of Nizwa, Nizwa, Sultanate of Oman}

\author{Ghulam Fatima}
\email{ghulamfatima.math@gmail.com}\affiliation{Department of
Physics, Zhejiang Normal University, Jinhua 321004, People's
Republic of China}

\begin{abstract}
A gravastar comprises three distinct sections: the interior zone, the middle shell, and its outer region. By considering a specific extended teleparallel gravity model that incorporates conformal Killing vectors and provides the field equations. We observe that the interior part exhibits a repellent force acting on the shell. This is based on the assumption that pressure is analogous to negative energy density. The middle shell consists of ultrarelativistic plasma and pressure, which is directly proportional to the matter density and counteracts the repellent force exerted by the inner zone. In the outer zone, we compute the precise solution in a vacuum and then connect these spacetimes using junction conditions to investigate stability limits. We aim to investigate the influence of dark energy models on the stable characteristics of gravastar configurations. It is worth noting that the phantom field exhibits the highest stable configurations for all physically viable selections of physical parameters. We additionally investigate the influence of physical parameters on the correct length, entropy, and energy of the gravastar.

\end{abstract}

\maketitle

\date{\today}

\section{Introduction}\label{sec:1}

The remarkable astrophysical conjecture like gravitational collapse and the exploration of stellar structures have gained the attention of several researchers. Neutron stars, white dwarfs and black holes (BHs) are novel examples of massive as well as dense compact stellar objects that are formed as an outcome of gravitational collapse. In recent years, many authors have suggested that the densest stellar models other than BHs can be produced by the gravitational collapse of a massive star. For this purpose, Mazur and Mottola \cite{1} by incorporating the modified notion of Bose-Einstein condensation in the gravitational field, presented a novel concept of a collapsing celestial body characterized as the gravastar. It is believed that such object coincides with the theoretical conditions for stable stellar development and provides an answer to the issues with classical BHs. In this hypothesis, it is expected that the quantum vacuum oscillations play a major part in the collapsing phenomenon. A phase transformation appears that results in the creation of a repellent de Sitter center, which keeps the collapsing candidate balanced and stops the singularity and horizon from being produced \cite{2}. This conversion occurs very near to the limit $\frac{2m}{r}=1$, rendering it exceedingly difficult for an outer observer to distinguish between the gravastar and an original BH.

The construction of gravastar structure is specified by three modes in which the inner era ($r\geq 0, r< \mathcal{R}_{1}$) is based on the isotropic de Sitter center possessing the $p=-\rho$ equation of state (EoS). The outer vacuum era ($r> \mathcal{R}_{2}$) is displayed through the Schwarzschild metric having $\rho=0=p$. The inner and outer modes are isolated by a thin shell ($\mathcal{R}_{1}<r< \mathcal{R}_{2}$) of stiff substance comprising the $\rho=p$ EoS possessing $\mathcal{R}_{1}$ as the inner and $\mathcal{R}_{2}$ as the outer radius of gravastar. A huge investigation on gravastar geometry has been accomplished after the demonstration of Mazur and Mottola. The stability of gravastar geometry via different EoSs is examined by Visser and Wiltshire \cite{3}. Carter \cite{4} computed the analytical solutions of gravastar and checked the consequences of EoSs on distinct modes of gravastar structure. By taking into account the Born-Infeld phantom in lieu of de Sitter line element, Bili\'{c} et al. \cite{5} evaluated the gravastar structures and presented that their outcomes can display the dense dark scenario. Horvat and Iliji\'{c} \cite {6} inspected the energy constraints inside the shell of gravastar and analyzed the stable geometry via different techniques. The study of thin-shell gravastars developed from inner de Sitter and outer various BH geometries is presented in  \cite{fa7}-\cite{fa12}.

Chirenti and Rezzola \cite{7} performed the stability against axial perturbations and found stable structures of gravastar. They also presented the differences of gravastar geometry from BHs. Rocha et al. \cite{8} constructed the models of prototype gravastars and displayed their differences from BHs. Harko et al. \cite{9} considered the accretion disks about slowly rotating gravastars and observed that the thermodynamic as well as electromagnetic attributes of the disks are distinct. They also exhibited that the gravastars yield a less efficient formalism for changing mass to radiation as compared to BHs. In the presence of nonlinear electrodynamics, Lobo and Arellano \cite{10} formed several models of gravastar and examined certain specific features of their structures. Similarly, with the effect of the electric field, Horvat et al. \cite{11} scrutinized the stability of the distinct zones of gravastar. Turimov et al. \cite{12} discussed the outcomes of magnetic field on the structure of gravastar and elaborated the analytical results in the presence of rotation. Javed and Lin \cite{13} explored the gravastar geometry by inducing the quintessence and cloud of strings.  The detailed study of thin-shell developing from different inner and outer manifold as studied in \cite{fk1}-\cite{fk6}.

It is considered that the occurrence of gravitational theories is due to the mysteries behind the cosmological expansion. In such frameworks, the matter and geometry coupling provides various speculations such as $f(R, T)$ theory \cite{14}, $f(R, T, R_{\alpha\beta}T^{\alpha\beta})$ theory \cite{15}, $f(\mathcal{G}, T)$ \cite{16} and $f(\mathcal{T}, T)$ theory \cite{17} having $R$ as the curvature scalar, $T$ exhibits the trace part of the stress energy tensor, $\mathcal{G}$ indicates the Gauss-Bonnet scalar and $\mathcal{T}$ manifests the torsion scalar. The discussion about the different modes of gravastar geometry has intrigued researchers to scrutinize the impact of different gravitational speculations on the structures of gravastar. In geometry-matter coupled $f(R,T)$ scenario, Das et al. \cite{19} determined the models of gravastar and explored its attributes graphically by incorporating several EoSs. In the light of $f(\mathcal{G},T)$ conjecture, Shamir and Ahmad \cite{20} derived the singularity-free solutions of gravastar and obtained the expressions of various physical factors. Sharif and Waseem \cite{21} observed the consequences of the Kuchowicz ansatz on the distinct zones of gravastar in $f(R,T)$ gravity.  In the light of $f(R,T^{2})$ background, Sharif and Naz \cite{25,26} determined the results of the absence as well as the appearance of electric charge on the geometrical attributes of gravastar. In $f(\mathcal{T}, T)$ theory, Ghosh et al. \cite{26a} computed the analytical and singularity free results of the gravastar yielding various physically viable characteristics.

For massive stellar objects, an inherited symmetry having a set of conformal Killing vectors (CKVs) is essential to establish a naturally coherent relationship between the elements of matter and geometry through field equations. These vectors are implemented to provide analytical results of the field equations more accurately as compared to the earlier existing methods. By incorporating such vectors, the giant system consists of nonlinear partial differential equations can be turned into ordinary differential equations. Implementing these KVs, Usmani et al. \cite{27} inspected the various aspects of gravastar by taking into account the electric field and evaluated the solutions for distinct modes of gravastar.
Sharif and Waseem \cite{28} scrutinized the significance of the charge factor on the modes of gravastar in the light of $f(R,T)$ scenario by utilizing the CKVs. Bhar and Rej \cite{29} manifested the charged geometry of gravastar with CKV in the light of $f(\mathcal{T})$ conjecture. Sharif et al. \cite{30,31} investigated the geometry of gravastar without charge and with charge by inducing the CKVs in $f(R,T^{2})$ gravity. Ghosh et al. \cite{gh7} explored the physical characteristics of gravastars solutions in $f(T)$ gravity.
The study of gravastars solutions in the framework of branworld gravity is presented in \cite{gh6} and \cite{gh5}. Bhar \cite{gh1} investigated gravastar remedies in the de Rham–Gabadadze–Tolley's large gravity is similar to that of massive gravity; explore many physical attributes.  In the context of f(Q) gravity, we build the non-singular solutions for charged gravastar \cite{gh2, gh3}. Also, the study of charged gravastars in Rastall gravity is presented in \cite{gh4}.

Like other matter-curvature coupled speculations, the $f(\mathcal{T}, T)$ gravity has also gained much attention of researchers in both astrophysical and cosmological scenarios. In this work, we establish the singularity-free results of charged gravastar in the context of $f(\mathcal{T}, T)$ theory with CKV. The paper is organized in such a way that the next section presents the basics of the $f(\mathcal{T}, T)$ scenario. Section III displays the physical exposure of charged gravastars along with the effects of CKVs. Section IV is devoted to matching the exact inner and outer solutions at the intermediate shell to establish the geometry of the gravastar. Then, we scrutinize the impact of phantomlike EoS on the stability of the gravastar model as presented in Section V. The structural attributes like the proper length of the intermediate thin shell, energy contents, and entropy are observed in Section VI. The findings of our work are manifested in the last section.

\section{ $f(\mathcal{T},T)$ Theory} \label{Sec:2}

To explore the $f(\mathcal{T},T)$ theory, we choose the following metric
\begin{equation}\label{r1}
ds^2=g_{\beta \xi}dx^{\beta}dx^{\xi}=\eta_{ij}\theta^{i}\theta^{j},
\end{equation}
with
\begin{equation} \label{r2}
dx^{\beta}=e_{i}{}^{\beta}\theta^{i};\;\; \theta^{i}=e^{i}{}_{\beta}dx^{\beta},
\end{equation}
where Minkowski metric is described by $\eta_{ij}$ with $diag(-,+,+,+)$. Further, the expression $e^{i}{}_{\beta}$ meets the following relations
\begin{equation}\label{r3}
e_{i}{}^{\beta}e^{i}{}_{\xi}=\delta^{\beta}_{\xi},\;\; e_{\beta}{}^{\xi}e^{\beta}{}_{j}=\delta^{\xi}_{j}.
\end{equation}
As we are working with the torsional based theory, so the Levi-Civita connection is provided as
\begin{equation}\label{r4}
\dot{\Gamma}^{\varrho}{}_{\beta\xi}=\frac{1}{2}g^{\varrho \sigma}(\partial_{\xi}g_{\sigma \beta}+\partial_{\beta}g_{\sigma \xi}-\partial_{\sigma}g_{\beta \xi}).
\end{equation}
The Weitzenb\"{o}ck connection or relation is defined as
\begin{equation}\label{r5}
\Gamma^{\gamma}_{\beta\xi}=e_{i}{}^{\gamma}\partial_{\beta}e^{i}{}_{\xi}=-e^{i}{}_{\xi}\partial_{\xi}e_{i}{}^{\gamma}.
\end{equation}
Within the scope of Weitzenb\"{o}ck connection by Eq. (\ref{r5}), the torsion becomes
\begin{equation}\label{r6}
\mathcal{T}^{\gamma}_{\beta\xi}=\Gamma^{\gamma}_{\beta\xi}-\Gamma^{\gamma}_{\xi\beta}.
\end{equation}
Further, the contorsion tensor is evaluated by
\begin{equation}\label{r7}
K^{\gamma}_{\beta\xi}=\Gamma^{\gamma}_{\beta\xi}-\dot{\Gamma}^{\gamma}_{\beta\xi}=\frac{1}{2}(\mathcal{T}_{\beta}{}^{\gamma}{}_{\xi}
+\mathcal{T}_{\xi}{}^{\gamma}{}_{\beta}-\mathcal{T}^{\gamma}{}_{\beta\xi}).
\end{equation}
The Eq. (\ref{r7}) can be redefined as
\begin{equation}\label{r8}
K^{\beta\xi}_{\gamma}=-\frac{1}{2}(\mathcal{T}^{\beta\xi}{}_{\gamma}-\mathcal{T}^{\xi\beta}{}_{\gamma}+\mathcal{T}_{\gamma}{}^{\beta\xi}).
\end{equation}
Th super-potential on combining torsion and contorsion is computed as
\begin{equation}\label{r9}
S_{\gamma}{}^{\beta\xi}=\frac{1}{2}(K^{\beta\xi}{}_{\gamma}+\delta^{\beta}_{\gamma} \mathcal{T}^{\sigma\xi}{}_{\sigma}-\delta^{\xi}_{\gamma} \mathcal{T}^{\sigma\beta}{}_{\sigma}).
\end{equation}
The Lagrangian density for torsion $\mathcal{T}$ expression is given by
\begin{equation}\label{r10}
\mathcal{T}=\mathcal{T}^{\gamma}{}_{\beta\xi}S_{\gamma}{}^{\beta\xi}.
\end{equation}
The Riemann tensor is
\begin{equation}\label{r11}
R_{\gamma}{}^{\beta\varrho\xi}=K^{\gamma}_{\beta\varrho;\xi}-K^{\gamma}_{\beta\xi;\varrho}+K^{\gamma}_{\sigma\xi}K^{\sigma}_{\beta\varrho}
-K^{\gamma}_{\sigma\varrho}K^{\sigma}_{\beta\xi}.
\end{equation}
From the above equation, one can evaluate the Ricci scalar as
\begin{equation}\label{r12}
R=-\mathcal{T}-2D_{\beta} \mathcal{T}^{\xi\beta}_{\xi},
\end{equation}
where $D_{\beta}$ is a covariant derivative. A standard action \cite{gjim1,hjim2} having matter coupling for extended teleparallel gravitational theory is provided as
\begin{equation} \label{1}
s=\int dx^{4} e\lbrace \frac{1}{2k^{2}} f(\mathcal{T},T)+\mathcal{L}_{(M)}\rbrace,
\end{equation}
where $e= det\left(e^{i}{}_{\beta}\right) =\sqrt{-g}$ and $k^{2}=8\pi G$. $\mathcal{L}_{(M)}$ provides the Lagrangian density. The field equations from the Eq. (\ref{1}) are calculated as
\begin{eqnarray} \label{rr2}
\frac{8\pi G}{2}T^Y_X &&=S_{\beta}^{\,\, \xi\varrho}\left(f_{\mathcal{T}\mathcal{T}}\partial_\varrho \mathcal{T}+f_{\mathcal{T}T}\partial_\varrho T\right)+\left[e^{-1} e^i_{\beta}\partial_\varrho(ee^\gamma_i S_\gamma^{\,\, \xi\varrho}\right.\nonumber\\&&\left.+\mathcal{T}^\gamma_{\,\, \gamma\beta}S_\gamma^{\,\, \xi\gamma}\right]f_\mathcal{T}+\frac{1}{4}\delta^\xi_\beta f+\frac{1}{2}f_{T}(T^Y_X+p_{t}\delta^\xi_\beta),
\end{eqnarray}
where $f_{\mathcal{T}T}=\frac{\partial^{2} f}{\partial \mathcal{T}\partial T}$ and $f_{\mathcal{T}}=\frac{\partial f}{\partial \mathcal{T}}$, $f_{T}=\frac{\partial f}{\partial T}$. On plugging Eq. (\ref{r11}) and Eq. (\ref{r12}) in Eq. (\ref{rr2}) the field equations become
\begin{equation}\label{2r1}
\left( R_{\beta\xi}-\frac{1}{2}g_{\beta\xi}R\right)f_\mathcal{T}+\frac{1}{2}g_{\beta\xi}(f-f_\mathcal{T}\mathcal{T})+2 S{\xi\beta}^{\,\,\sigma}(f_{\mathcal{T}\mathcal{T}}\bigtriangledown_{\sigma}\mathcal{T}+f_{\mathcal{T}\mathcal{T}}\bigtriangledown_{\sigma}\mathcal{T})=
(8\pi G-f_{\mathcal{T}})\mathcal{T}_{\beta\xi}-f_{\mathcal{T}}g_{\beta\xi}p_{t}.
\end{equation}
In order to complete this analysis, the following line element is considered
\begin{equation}\label{3}
d{s}^2=-e^{X(r)}dt^2+e^{Y(r)}d{r}^2+r^{2} d\theta^{2}+r^{2}\sin^{2}\theta d\phi^{2},
\end{equation}
For the isotropic fluid source, the energy-momentum tensor reads as
\begin{equation}\label{eq:14a}
T_{ij}=diag(e^{X}\rho, e^{Y}p,r^{2} p, r^2 p sin^{2}\theta).
\end{equation}
Being modification in matter part, the electromagnetic stress-energy
tensor $E_{i \upsilon}$ is defined by
\begin{equation}\label{eq:14b}
E_{i j}=2(F_{i \zeta}F_{j \zeta}-\frac{1}{4}g_{i j}F_{\zeta
\chi}F^{\zeta \chi}),
\end{equation}
where
\begin{equation*}
F_{i j}=\mathcal{A}_{i, j}-\mathcal{A}_{j, i}.
\end{equation*}
Further, a tensor, i.e., $F_{i j}$ is expressed in the following form
\begin{eqnarray}
F_{i j,\zeta}+F_{\zeta i,j}+F_{j \zeta,i}=0,\label{22}\\
(\sqrt{-g}F^{i j})_{,j}=\frac{1}{2}\sqrt{-g} j^{i}.\label{23}
\end{eqnarray}
Now, the Eq. (\ref{eq:14b}) gives us the following relation of
electric field ($E$).
\begin{equation}\label{24}
E(r)=\frac{1}{2r^2}e^{X(r)+Y(r)}\int_{0}^{r}\sigma(r)e^{Y(r)}r^{2}dr=\frac{q(r)}{r^{2}}.
\end{equation}
In the above equation, $q(r)$ and $\delta$ define the total charge as well as the charge density for the system, respectively. Finally, the diagonal tetrad is computed by
\begin{equation}\label{5}
e_{\gamma}^{Y} =\Big(e^{\frac{X(r)}{2}},e^{\frac{Y(r)}{2}},r,r\sin\theta\Big).
\end{equation}
The determinant of $e_{\gamma}^{Y}$ yields $e=e^{X(r)+Y(r)}r^{2}\sin\theta$.
The torsion $\mathcal{T}$ becomes
\begin{equation}
\mathcal{T}(r)=\frac{2 e^{-Y (r)}}{r} \left(X'(r)+\frac{1}{r}\right).\label{7}
\end{equation}
As, we are working with the diagonal tetrad, a suitable choice of model for $f(\mathcal{T},T)$ theory  \cite{gjim1,hjim2} is presented as:
\begin{equation}
f(\mathcal{T},T)=\alpha\mathcal{T}(r)+c_{1} T+c_{2}\label{8},
\end{equation}
where $\alpha,\;c_{1}$ and $c_{2}$ are functional parameters. By using Eq. (\ref{3}), Eq. (\ref{5}), and Eqs. (\ref{7}-\ref{8}) in Eq. (\ref{2r1}), one can get the $f(\mathcal{T},T)$ field equations as
\begingroup
\small
\begin{eqnarray}
&&\hspace{-1.6cm}8 \pi\rho =-\frac{e^{-Y (r)}}{4 (c_{1}-1)(c_{1} +2) r^2}\bigg(2 \alpha  c_{1} -4 \alpha +\alpha  c_{1}  r^2 Y '(r) X '(r)-2 \alpha \nonumber\\&&\hspace{-1cm} \times c_{1}  r^2 X ''(r)-\alpha  c_{1}  r^2 X '(r)^2+c_{1}  r^2 c_{2}  e^{Y (r)}+2 r^2 c_{2}  e^{Y (r)}-\alpha  c_{1}  r \nonumber\\&&\hspace{-1cm} \times Y '(r)-2 \alpha  c_{1}  e^{Y (r)}-3 \alpha  c_{1}  r X '(r)+4 \alpha  r Y '(r)+4 \alpha  e^{Y (r)}\bigg),~~~~\label{9}\\
&&\hspace{-1.6cm}8 \pi p=\frac{e^{-Y (r)}}{4 (c_{1} -1) (c_{1} +2) r^2}\bigg(2 \alpha  c_{1} -4 \alpha +\alpha  c_{1}  r^2 Y '(r) X '(r)-2 \alpha \nonumber\\&&\hspace{-1cm} \times c_{1}  r^2 X ''(r)-\alpha  c_{1}  r^2 X '(r)^2+c_{1}  r^2 c_{2}  e^{Y (r)}+2 r^2 c_{2}  e^{Y (r)}+3 \alpha  c_{1} \nonumber\\&&\hspace{-1cm} \times r Y '(r)-2 \alpha  c_{1}  e^{Y (r)}+\alpha  c_{1}  r X '(r)+4 \alpha  e^{Y (r)}-4 \alpha  r X '(r)\bigg),~~~\label{10}\\
&&\hspace{-1.6cm}8 \pi p=\frac{e^{-Y (r)}}{4 (c_{1} -1) (c_{1} +2) r^2}\bigg(\alpha  r^2 Y '(r) X '(r)-2 \alpha  c_{1}-2 \alpha  r^2 X ''(r)\nonumber\\&&\hspace{-1cm} -\alpha  r^2 X '(r)^2+c_{1}  r^2 c_{2}  e^{Y (r)}+2 r^2 c_{2}  e^{Y (r)}+\alpha  c_{1}  r Y '(r)+2 \alpha  c_{1} \nonumber\\&&\hspace{-1cm} \times e^{Y (r)}-\alpha  c_{1}  r X '(r)+2 \alpha  r Y '(r)-2 \alpha  r X '(r)\bigg).\label{11}
\end{eqnarray}
\endgroup

An efficient way to use the field equations to follow the natural link between matter and geometry is to employ inheritance symmetry. Typically, inheritance symmetry is used to measure the symmetry resulting from the conformal Killing vectors. Moreover, we employ the vector field $h$, which may be represented as follows, to further utilize the idea of conformal symmetry
\begin{equation}\label{12}
\mathfrak{L}_{Z} g_{XY}=h(r)g_{XY},
\end{equation}
where $\mathfrak{L}$ represents the Lie derivative and vector field is denoted by $h(r)$. Using
Eq.(\ref{3}) in Eq.(\ref{12}), one can get following relations
\begin{eqnarray*}
Z^{1}X^{'}(r)=h(r),\;\;\;\;\;\;
Z^{1}=\frac{rh(r)}{2},\;\;\;\;\;\;
Z^{1}Z^{'}(r)+2 Z^{1}_{,1}=h(r).
\end{eqnarray*}
The solutions of this system within the scope of Eq. (\ref{3}) yield the following expressions
\begin{eqnarray}
e^{X (r)}=K_{1}r^2, \;\;\;\;\;\;\;\;e^{Y
(r)}=\frac{K_{2}}{h^{2}(r)},\label{13}
\end{eqnarray}
where, $K_1$ and $K_2$ are integration constants. Now, inserting the Eq.(\ref{13}) in Eqs. (\ref{9}-\ref{11}), we attain the modified field equations as:
\begin{eqnarray}
8 \pi\rho &=&\frac{4 \alpha  (c_{1} +1) h^2(r)+2 \alpha  (c_{1} +4) r h(r) h'(r)+K_2 \left(2 \alpha  (c_{1} -2)-(c_{1} +2) r^2 c_{2} \right)}{4 (c_{1} -1) (c_{1} +2) K_2 r^2},\label{14}\\
8 \pi p&=&\frac{4 \alpha  (c_{1} -3) h^2(r)-10 \alpha  c_{1}  r h(r) h'(r)+K_2 \left((c_{1} +2) r^2 c_{2} -2 \alpha  (c_{1} -2)\right)}{4 (c_{1} -1) (c_{1} +2) K_2 r^2},\label{15}\\
8 \pi p&=&\frac{-4 \alpha  (c_{1} +1) h^2(r)-2 \alpha  (c_{1} +4) r h(r) h'(r)+K_2 \left(2 \alpha  c_{1} +(c_{1} +2) r^2 c_{2} \right)}{4 (c_{1} -1) (c_{1} +2) K_2 r^2}.\label{16}
\end{eqnarray}

\section{Charged gravastar with Conformal Motion}

Here, we shall deal with the new charged gravastar solution under conformal symmetry. On using Eq. (\ref{13}) into Eqs.
(\ref{14}-\ref{16}) within Eq. (\ref{24}), one can get the following field equations
\begin{eqnarray}
8 \pi\rho +E^{2} &&= \frac{4 \alpha  (c_{1} +1) h^2(r)+2 \alpha  (c_{1} +4) r h(r) h'(r)+K_2 \left(2 \alpha  (c_{1} -2)-(c_{1} +2) r^2 c_{2} \right)}{4 (c_{1} -1) (c_{1} +2) K_2 r^2},\label{eq:27}\\
8 \pi p -E^{2} &&= \frac{4 \alpha  (c_{1} -3) h^2(r)-10 \alpha  c_{1}  r h(r) h'(r)+K_2 \left((c_{1} +2) r^2 c_{2} -2 \alpha  (c_{1} -2)\right)}{4 (c_{1} -1) (c_{1} +2) K_2 r^2},\label{eq:28}\\
8 \pi p +E^{2} &&= \frac{-4 \alpha  (c_{1} +1) h^2(r)-2 \alpha  (c_{1} +4) r h(r) h'(r)+K_2 \left(2 \alpha  c_{1} +(c_{1} +2) r^2 c_{2} \right)}{4 (c_{1} -1) (c_{1} +2) K_2 r^2}.\\\label{eq:29}
\end{eqnarray}
The expressions of physical factors are obtained simultaneously by solving the above field equations as
\begin{eqnarray}
 \rho &&= -\frac{K_2 \left(2 \alpha +(c_{1} +2) r^2 c_{2} \right)+2 \alpha  (c_{1} -6) r h (r) h '(r)-8 \alpha  c_{1}  h (r)^2}{32 \pi  \left(c_{1} ^2+c_{1} -2\right) K_2 r^2},\label{eq:30}\\
 p &&= \frac{K_2 \left(2 \alpha +(c_{1} +2) r^2 c_{2} \right)-2 \alpha  (3 c_{1} +2) r h (r) h '(r)-8 \alpha  h (r)^2}{32 \pi  \left(c_{1} ^2+c_{1} -2\right) K_2 r^2},\label{eq:31}\\
    E^{2} &&= \frac{\alpha  \left(K_2+2 r h (r) h '(r)-2 h (r)^2\right)}{2 (c_{1} +2) K_2 r^2}.\\\label{eq:32}
\end{eqnarray}

\subsection{Interior mode of charged gravastar}

From Eqs. (\ref{eq:30}) and (\ref{eq:31}), the relation between the physical terms and the metric functions is given by
\begin{eqnarray}\label{eq:33}
\rho+p=\frac{\alpha  h (r) \left(h (r)-r h '(r)\right)}{4 \pi  (c_{1} +2) K_2 r^2}.
\end{eqnarray}
Employing the condition $\rho+p=0$, we obtain the exact form for
$h (r)$ through Eq. (\ref{eq:33}), which leads to
\begin{eqnarray}\label{eq:34}
h (r)=rh_{0}  ,\;\;\;\;\;h (r)=0,
\end{eqnarray}
where $h_{0}$ acts as an integration constant. Here, $h (r)=0$ does not provide a valid solution. On inserting the expressions of $h (r)$, the physical variables become
\begin{eqnarray}
 \rho&&= -\frac{K_2 \left(2 \alpha +(c_{1} +2) r^2 c_{2} \right)-6 \alpha  (c_{1} +2) r^2 h_{0}^2}{32 \pi  \left(c_{1} ^2+c_{1} -2\right) K_2 r^2}=-p,\label{eq:35}\\
  E^{2}&& =\frac{\alpha  \left(K_2+2 r h (r) h '(r)-2 h (r)^2\right)}{2 (c_{1} +2) K_2 r^2},\label{eq:36}\\
\sigma &&=\frac{h_{0} \sqrt{\frac{\alpha }{(c_{1} +2) r^2}}}{4 \sqrt{2} \pi  K_2 }. \label{eq:37}
\end{eqnarray}

The corresponding forms of lapse functions yield
\begin{equation}\label{20}
e^{a}=K_{1}r^{2}, \quad e^{-b}=\frac{r^2 h_0^2}{K_{2}}.
\end{equation}
The active gravitational mass $M(r)$ through
Eq. (\ref{eq:30}) is determined as
\begin{eqnarray}
M(r)&&=4 \pi  \left(\frac{\alpha  r^3 h_{0}^2}{8 \pi
\left(c_{1} ^2+c_{1} -2\right) K_2}+\frac{\alpha  c_{1}  r^3 \text{$h
$0}^2}{16 \pi  \left(c_{1} ^2+c_{1} -2\right) K_2}-\frac{r^3 c_{2}
}{48 \pi  \left(c_{1} ^2+c_{1} -2\right)}-\frac{c_{1}  r^3 c_{2}
}{96 \pi  \left(c_{1} ^2+c_{1}
-2\right)}\right.\\\nonumber&-&\left.\frac{\alpha r}{16 \pi
\left(c_{1} ^2+c_{1} -2\right)}+\frac{\alpha  r}{16 \pi (c_{1}
+2)}\right).\label{eq:38}
\end{eqnarray}

\subsection{Charged gravastar shell with conformal motion}

Here, we are going to compute the exact physical expressions
in the light of $\rho=p$ EoS by implementing Eq.
(\ref{eq:30}) and Eq. (\ref{eq:31}). $\rho=p$ is given by
\begin{eqnarray}
&&{}-\frac{K_2 \left(2 \alpha +(c_{1} +2) r^2 c_{2} \right)+2 \alpha  (c_{1} -6) r h (r) h '(r)-8 \alpha  c_{1}  h (r)^2}{32 \pi  \left(c_{1} ^2+c_{1} -2\right) K_2 r^2}\nonumber\\&&=\frac{K_2 \left(2 \alpha +(c_{1} +2) r^2 c_{2} \right)-2 \alpha  (3 c_{1} +2) r h (r) h '(r)-8 \alpha  h (r)^2}{32 \pi  \left(c_{1} ^2+c_{1} -2\right) K_2 r^2}.\label{eq:39}
\end{eqnarray}
Embedding $\rho=p$, the expression for $h
(r)$ via Eq. (\ref{eq:39}) is obtained as
\begin{eqnarray}\label{eq:40}
h=\frac{\sqrt{r^{-\frac{4 (c_{1} +1)}{c_{1} +4}} \left(6 \alpha  (c_{1} +1) h_{1}+3 \alpha  K_2 r^{\frac{4 (c_{1} +1)}{c_{1} +4}}+(c_{1} +1) K_2 c_{2}  r^{\frac{6 (c_{1} +2)}{c_{1} +4}}\right)}}{\sqrt{6} \sqrt{\alpha  (c_{1} +1)}}.
\end{eqnarray}
with $h_{1}$ as an integration constant.
$h (r)=0$ does not yield a valid solution. Now, substituting the forms
of $h (r)$, the physical variables take the forms
\begin{eqnarray}
\rho &&= \frac{r^{-\frac{6 (c_{1} +2)}{c_{1} +4}} \left(6 \alpha  (c_{1} +1) (c_{1} +2) h_{1}+\alpha  (c_{1} +4) K_2 r^{\frac{4 (c_{1} +1)}{c_{1} +4}}\right)}{16 \pi  (c_{1} +1) (c_{1} +2) (c_{1} +4) K_2}=p,\label{eq:41}\\
E^{2}&&=\frac{r^{-\frac{6 (c_{1} +2)}{c_{1} +4}} \left(\alpha  c_{1}  (c_{1} +4) K_2 r^{\frac{4 (c_{1} +1)}{c_{1} +4}}-6 \alpha  (c_{1} +1) (c_{1} +2) h_{1}\right)}{2 (c_{1} +1) (c_{1} +2) (c_{1} +4) K_2},\label{eq:42}\\
\sigma&& =\frac{\alpha  r^{\frac{12}{c_{1} +4}-7} \left(6 (c_{1} -2) (c_{1} +1) (c_{1} +2) h_{1}+c_{1}  (c_{1} +4)^2 K_2 r^{\frac{4 (c_{1} +1)}{c_{1} +4}}\right) \sqrt{\frac{3}{c_{1} +1}+\frac{6 h_{1} r^{-\frac{4 (c_{1} +1)}{c_{1} +4}}}{K_2}+\frac{r^2 c_{2} }{\alpha }}}{8 \sqrt{3} \pi  (c_{1} +1) (c_{1} +2) (c_{1} +4)^2 K_2 \sqrt{\frac{\alpha  r^{-\frac{6 (c_{1} +2)}{c_{1} +4}} \left(c_{1}  (c_{1} +4) K_2 r^{\frac{4 (c_{1} +1)}{c_{1} +4}}-6 (c_{1} +1) (c_{1} +2) h_{1}\right)}{(c_{1} +1) (c_{1} +2) (c_{1} +4) K_2}}}.\label{eq:43}
\end{eqnarray}

The associated lapse functions are
\begin{equation}\label{44}
e^{a}=K_{1}r^{2}, \quad e^{-b}=\frac{1}{2 c_{1} +2}+\frac{h_{1}
r^{-\frac{4 (c_{1} +1)}{c_{1} +4}}}{K_2}+\frac{r^2 c_{2} }{6 \alpha
}.
\end{equation}

The active gravitational mass using Eq. (\ref{eq:30}) becomes
\begin{eqnarray}\label{eq:45}
M(r)=\frac{4 \pi  \alpha  r}{16 \pi  c_{1} +32 \pi }.
\end{eqnarray}

\subsection{External mode of the charged gravastar with conformal vectors}

We shall evaluate the analytic forms of physical
factors using $p=\omega\rho$ EoS possessing $\omega=0$ through Eq. (\ref{eq:30}) and Eq. (\ref{eq:31}). Considering the
EoS, the expression for $h (r)$ through Eq.
(\ref{eq:39}) is obtained as
\begin{eqnarray}\label{eq:46}
h(r)=\frac{((3 c_{1} +2) r)^{-\frac{4}{3 c_{1} +2}} \sqrt{12 \alpha  h_{2}+K_2 ((3 c_{1} +2) r)^{\frac{8}{3 c_{1} +2}} \left(3 \alpha +2 r^2 c_{2} \right)}}{2 \sqrt{3} \sqrt{\alpha }}.
\end{eqnarray}
in which $h_{2}$ exhibits the constant of integration. $h (r)=0$ does not yield a consistent solution. On inserting $h (r)$, the forms of physical factors become
\begin{eqnarray}
E^{2}&&=\frac{\alpha  \left(\frac{1}{c_{1} +2}-\frac{12 h_{2} r ((3 c_{1} +2) r)^{-\frac{8}{3 c_{1} +2}-1}}{K_2}\right)}{4 r^2},\label{eq:47}\\
\sigma &&=\frac{\alpha  ((3 c_{1} +2) r)^{-\frac{8}{3 c_{1} +2}} \left((3 c_{1} +2)^2 K_2 ((3 c_{1} +2) r)^{\frac{8}{3 c_{1} +2}}-12 (c_{1} +2) (3 c_{1} -2) h_{2}\right) \sqrt{\frac{1}{\frac{h_{2} ((3 c_{1} +2) r)^{-\frac{8}{3 c_{1} +2}}}{K_2}+\frac{r^2 c_{2} }{6 \alpha }+\frac{1}{4}}}}{8 \pi  (c_{1} +2) (3 c_{1} +2)^2 K_2 r^3 \sqrt{\frac{\alpha  \left(\frac{1}{c_{1} +2}-\frac{12 h_{2} r ((3 c_{1} +2) r)^{-\frac{8}{3 c_{1} +2}-1}}{K_2}\right)}{r^2}}},\label{eq:48}
\end{eqnarray}
The lapse functions in this case are
\begin{equation}\label{49}
e^{a}=K_{1}r^{2}, \quad e^{-b}=\frac{1}{\frac{h_{2} ((3 c_{1} +2) r)^{-\frac{8}{3 c_{1} +2}}}{K_2}+\frac{r^2 c_{2} }{6 \alpha }+\frac{1}{4}}.
\end{equation}

Now, we determine the Kretschmann scalar $(K_S)$ to exhibit that the external mode is a flat geometry in the following form
\begin{eqnarray}\label{eq:44}
K_S= R^{ijkl}R_{ijkl},
\end{eqnarray}
where $R$ indicates the Riemann tensor. The $K_S$ for
the external zone becomes,

\begin{eqnarray}\label{eq:44a}
K_S=\frac{\mathcal{J}_3 \left(144 \alpha ^2 (3 c_{1}  (15 c_{1}
+4)+52) h_{0}^2+\mathcal{J}_1+\mathcal{J}_2\right)}{432 \alpha
^4 (3 c_{1} +2)^2 K ^{4}_{2} r^4 ((3 c_{1} +2) r)^{\frac{32}{3 c_{1}
+2}}},
\end{eqnarray}
where
\begin{eqnarray}\nonumber
\mathcal{J}_1&=&(3 c_{1} +2)^2 K ^{2}_{2} ((3 c_{1} +2)
r)^{\frac{16}{3 c_{1} +2}} \left(45 \alpha ^2+52 r^4 c_{2} ^2+84
\alpha  r^2 c_{2} \right),\\\nonumber \mathcal{J}_2&=&24 \alpha  (3
c_{1} +2) h_{0}K_{2} ((3 c_{1} +2) r)^{\frac{8}{3 c_{1}
+2}} \left(\alpha  (45 c_{1} +6)+2 (21 c_{1} -10) r^2 c_{2}
\right),\\\nonumber \mathcal{J}_3&=&\left(12 \alpha
h_{0}+K_{2} ((3 c_{1} +2) r)^{\frac{8}{3 c_{1} +2}} \left(3
\alpha +2 r^2 c_{2} \right)\right)^2.
\end{eqnarray}
Notice that $K_S\rightarrow 0$ as
$r\rightarrow \infty$ which interprets that the
external structure is an asymptotically flat geometry.

\subsection{Boundary constraints}

Boundary conditions are crucial in calculating the expressions of various parameters that arise within the system. Now, by comparing the inner and thin shell regions at the surface $r = r_1$, we derive the factor A and investigate its dependence on charge $q$. Additionally, by comparing the thin shell domain with the external area at $r = r_2$, we calculate the thickness ($r_2 - r_1$) of the thin shell as well as the range of the outside radius. The internal radius is taken into account as $R_1 = 10km$ \cite{44}.

\begin{itemize}

\item  Comparing the inner and thin shell eras at $r = r_1 = 10 km$, we get:
\begin{eqnarray}\label{eq:44b}
K_{1}r^{2}_1=K_{1}r^{2}_1,
\end{eqnarray}
for $g_{tt}$ component and also for $g_{rr}$ component, we have
\begin{eqnarray}\label{eq:44c}
\frac{r^2 h_{0}^2}{K_{2}}=\frac{r^2_1 r_1^{-\frac{4
(c_{1} +1)}{c_{1} +4}} \left(6 \alpha  (c_{1} +1) h_{1}+3 \alpha
K_{2} r_1^{\frac{4 (c_{1} +1)}{c_{1} +4}}+(c_{1} +1) K_{2}
c_{2}  r_1^{\frac{6 (c_{1} +2)}{c_{1} +4}}\right)}{6 \alpha  (c_{1} +1)
K_{2}}.
\end{eqnarray}

\item Moreover, through the thin shell and the external region at $r = r_2$, we gain
\begin{eqnarray}\label{eq:44d}
K_{1} r_2^{2}=K_{1} r_2^{2},
\end{eqnarray}
for $g_{tt}$ component and also for $g_{rr}$  component, we have
\begin{eqnarray}\label{eq:44e}
\frac{r^2_2 r_2^{-\frac{4 (c_{1} +1)}{c_{1} +4}} \left(6 \alpha  (c_{1}
+1) h_{1}+3 \alpha K_{2} r_2^{\frac{4 (c_{1} +1)}{c_{1}
+4}}+(c_{1} +1) K_{2} c_{2}  r_2^{\frac{6 (c_{1} +2)}{c_{1}
+4}}\right)}{6 \alpha  (c_{1} +1) K_{2}}=\frac{h_{2} ((3
c_{1} +2) r_2)^{-\frac{8}{3 c_{1} +2}}}{K_{2}}+\frac{r^2_2 c_{2} }{6
\alpha }+\frac{1}{4}.
\end{eqnarray}

\end{itemize}

\section{Junction Conditions}\label{sec6}

Here, we match the inner and outer calculated solutions of
gravastars at the hypersurface. The manifolds mathematically can be
illustrated by
\begin{eqnarray}\label{1aa}
ds^{2}_\pm=\mathcal{F}_\pm(r_\pm)dt^{2}_\pm-\mathcal{F}_\pm(r_\pm)^{-1}dr^{2}_\pm-r^{2}_\pm
d\theta^{2}_\pm-r^{2}_\pm\sin^2{\theta}_\pm d\phi^{2}_\pm ,~~~
\end{eqnarray}
where the metric constituents of intrinsic and extrinsic manifolds
are computed with the help of field equations via boundary constraints.
The lapse function of external (+) manifold can be portrayed as
\begin{equation}\label{2aa}
\mathcal{F}_+(r_+)=\frac{K_{2} r_+^{\frac{4 (c_{1} +1)}{c_{1}
+4}} ((3 c_{1} +2) r_+)^{\frac{8}{3 c_{1} +2}} \left(\alpha  \left(6
r^2_+-3 (c_{1} +1)\right)+2 (c_{1} +1) \left(r^2_+-1\right) r^2_+
c_{2} \right)}{12 \alpha  (c_{1} +1) \left(r_+^{\frac{4 (c_{1}
+1)}{c_{1} +4}}-r^2 _+((3 c_{1} +2) r_+)^{\frac{8}{3 c_{1}
+2}}\right)}.
\end{equation}
While for interior, we have
\begin{equation}\label{2ab}
\mathcal{F}_-(r_-)=\frac{r_-^{\frac{4 (c_{1} +1)}{c_{1} +4}} \left(6
\alpha  (c_{1} +1) h_{0}^2-K_{2} \left(3 \alpha
+(c_{1} +1) r_-^2 c_{2} \right)\right)}{6 \alpha  (c_{1} +1)}.
\end{equation}

In the present study, the thin-shell enclosing WH geometry is created by means of a cut-and-paste method. Using this method, a unique regular manifold is produced, which can be expressed as $\mathcal{W}=\mathcal{W}^{-}\cup \mathcal{W}^{+}$. The resultant structure becomes non-singular when the shell radius is less than $r_h$. The components of the proposed manifolds and hypersurfaces have the following forms, according to the Israel formalism: $z^{\gamma}_\pm=(t_\pm,r_\pm,\theta_\pm,\phi_\pm)$ and $\eta^{i}=(\tau,\theta,\phi)$, respectively. Here $\tau$ specifies
the proper time at the hypersurface. These constituents are
connected with each other with the help of the following coordinate
transformation

\begin{eqnarray}\label{5aa}
g_{ij}=\frac{\partial
z^{\gamma}_\pm}{\partial\eta^{i}}\frac{\partial
z^{c_{1}}_\pm}{\partial\eta^{j}}g_{\gamma c_{1}}^\pm.
\end{eqnarray}

The parametric equation corresponding to the hypersurface is expressed by
\begin{eqnarray}\nonumber
\mathcal{H}: R(r,\tau)=r-y(\tau)=0.
\end{eqnarray}

The Lanczos equations at hypersurface, which are the reduced
forms of the Einstein field equations, are utilized to examine the matter
axioms displayed by
\begin{equation}\label{6aa}
S_{c_{1}}^{\alpha}=\frac{1}{8\pi}(\delta_{c_{1}}^{\alpha}
\zeta_{\gamma}^{\gamma}-\zeta_{c_{1}}^{\alpha}),
\end{equation}
where $\zeta_{\alpha c_{1}}=K^{+}_{\alpha c_{1}}-K^{-}_{\alpha c_{1}}$
and $K^{-}_{\alpha c_{1}}$ manifest the constituents of extrinsic
curvature. For perfect matter configuration, the stress-energy tensor
is described as
${S^{\alpha}}_{c_{1}}=diag(\mathfrak{S},\mathfrak{P},\mathfrak{P})$.
The surface density and pressure at shell are expressed by
$\mathfrak{S}$ and $\mathfrak{P}$, respectively. The extrinsic curvatures for inner and outer geometries are
\begin{equation}\label{7aa}
{K_{\alpha c_{1}}^{\pm}}= -n_{\mu}^\pm \left[\frac{\partial^2
z^{\mu}_\pm}{\partial \eta^{\alpha}
\eta^{c_{1}}}+\Gamma^{\mu}_{\lambda\nu}\left(\frac{\partial
z^{\lambda}_\pm}{\partial\eta^{\alpha}}\right)\left(\frac{\partial
z^{\nu}_\pm}{\partial\eta^{c_{1}}}\right)\right].
\end{equation}
The unit normals become
\begin{equation}\label{8aa}
n_{\pm}^{\mu}=\left(\frac{\dot{y}}{\mathcal{F}_\pm(y)},\sqrt{\dot{y}^2+\mathcal{F}_\pm(y)},0,0\right),
\end{equation}
in which the overdot indicates the differential associated with proper time.

Employing Lanczos equations, we obtain
\begin{eqnarray}\label{paa}
\mathfrak{S}&=&-\frac{[K^\theta_\theta]}{4\pi}=-\frac{1}{4\pi
y}\left\{\sqrt{\dot{y}^2+\mathcal{F}_+(y)}-\sqrt{\dot{y}^2+\mathcal{F}_-(y)}\right\},
\\\label{pbb}
\mathfrak{P}&=&\frac{[K^\theta_\theta]+[K_{\tau}^{\tau}]}{8\pi}=\frac{1}{8\pi
y}\left\{\frac{2\dot{y}^2+2y\ddot{y}
+2\mathcal{F}_+(y)+y\mathcal{F}'_+(y)}{\sqrt{\dot{y}^2+\mathcal{F}_+(y)}}-\frac{2\dot{y}^2+2y\ddot{y}
+2\mathcal{F}_-(y)+y
\mathcal{F}'_-(y)}{\sqrt{\dot{y}^2+\mathcal{F}_-(y)}}\right\}.
\end{eqnarray}

It is assumed that the thin shell of the generated geometry has no movement in a radial direction at equilibrium shell radius $y_0$. Hence, it is observed that the proper time
differential of shell radius disappears, i.e., $\dot{y_0}=0=\ddot{y_0}$.
\begin{eqnarray}\label{p1a}
\mathfrak{S}_0&=&-\frac{1}{4\pi
y_0}\left\{\sqrt{\mathcal{F}_+(y_0)}-\sqrt{\mathcal{F}_-(y_0)}\right\},
\\\label{p2a}
\mathfrak{P}_0&=&\frac{1}{8\pi
y_0}\left\{\frac{2\mathcal{F}_+(y_0)+y_0\mathcal{F}'_+(y_0)}
{\sqrt{\mathcal{F}_+(y_0)}}-\frac{2\mathcal{F}_-(y_0)+y_0
\mathcal{F}'_-(y_0)}{\sqrt{\mathcal{F}_-(y_0)}}\right\},
\end{eqnarray}
where at equilibrium position, density and pressure are presented by
$\mathfrak{S}(y_0)$ and $\mathfrak{P}(y_0)$, respectively.

\section{Stability via Linearized Radial Perturbation}\label{sec7}

We now going to discuss the stable state of a created
thin-shell about WH geometry via linearized radial
perturbation at $y=y_ 0$. We obtain the equation of motion of the shell
through Eq. (\ref{paa}) given by
\begin{equation}\label{13aa}
\Pi(y)+\dot{y}^2=0,
\end{equation}
having $\Pi(y)$ as effective potential function
illustrated by
\begin{eqnarray}\label{14aa}
\Pi(y)=-\frac{(\mathcal{F}_-(y)-\mathcal{F}_+(y))^2}{64 \pi ^2 y^2
\mathfrak{S} (y)^2}+\frac{1}{2}
(\mathcal{F}_-(y)+\mathcal{F}_+(y))-4 \pi ^2 y^2 \mathfrak{S} (y)^2.
\end{eqnarray}

It is interesting that the constituents of the stress-energy tensor
obey the energy conservation requirements given by
\begin{equation}\label{15aa}
\frac{d}{d\tau}(4\pi
y^2\mathfrak{S}(y))+\mathfrak{P}(y) \frac{d}{d\tau}(4\pi y^2)=0,
\end{equation}
which yields
\begin{equation}\label{16aa}
\mathfrak{S}'(y)=-\frac{\mathfrak{P}(\mathfrak{S}(y)))+2(\mathfrak{S}(y)}{y}.
\end{equation}

In order to expand the effective potential around the equilibrium shell radius to second-order factors for stable geometry, we implement the Taylor expansion given by
\begin{eqnarray}
\Pi(y)=\Pi(y_{0})+(y-y_{0})\Pi'(y_{0})+\frac{1}{2}
(y-y_{0})^2\Pi''(y_{0})+O[(y-y_{0})^3].~~~
\end{eqnarray}
The resulting geometry becomes stable for $\Pi(y_0)=0=\Pi'(y_0)$.
Thus, we obatin
\begin{equation}\label{17aa}
\Pi(y)=\frac{1}{2}(y-y_{0})^2\Pi''(y_{0}).
\end{equation}

To manifest the stability of thin-shell, the second differential form of the effective
potential at $y=y_0$ can be implemented as
\begin{itemize}
\item Unstable $\Rightarrow$\quad $\Pi''(y_{0})<0$,
\item Stable $\Rightarrow$\quad $\Pi''(y_{0})>0$,
\item Unpredictable $\Rightarrow$\quad  $\Pi''(y_{0})=0$.
\end{itemize}

\begin{eqnarray}\nonumber
\Pi''(y_{0})&=&\left(y_0 \mathfrak{S} (y_0) \left(y_0 \mathfrak{S}
(y_0) \left(16 \pi ^2 \mathcal{A}_2 y_0^2 \mathfrak{S}
(y_0)^2-\left(\mathcal{F}_-'(y_0)-\mathcal{F}_+'(y_0)\right)^2\right)-\mathcal{A}_4
\mathcal{F}_-(y_0)\right)\right.\\\label{sad}&+&\mathcal{F}_+(y_0)
\left(\mathcal{A}_3 y_0 \mathfrak{S} (y_0)-2 \mathcal{A}_5
\mathcal{F}_-(y_0)\right)+\left.\mathcal{A}_5
\mathcal{F}_-(y_0)^2+\mathcal{A}_1
\mathcal{F}_+(y_0)^2\right)\left(32 \pi ^2 y_0^4 \mathfrak{S}
(y_0)^4\right)^{-1},
\end{eqnarray}
where
\begin{eqnarray}\nonumber
\mathcal{A}_1&=&3 \mathfrak{S} (y_0) (2
\mathfrak{P}(y_0)+\mathfrak{S} (y_0))-4
(\mathfrak{P}(y_0)+\mathfrak{S} (y_0)) \left(-\mathfrak{S} (y_0)
\mathfrak{P}'(y_0)+3 \mathfrak{P}(y_0)+\mathfrak{S}
(y_0)\right),\\\nonumber
\mathcal{A}_2&=&\mathcal{F}_-''(y_0)+\mathcal{F}_+''(y_0)+16 \pi ^2
\left(\mathfrak{S} (y_0) (2 \mathfrak{P}(y_0)+\mathfrak{S} (y_0))-4
(\mathfrak{P}(y_0)+\mathfrak{S} (y_0)) \left(\mathfrak{S} (y_0)
\left(\mathfrak{P}'(y_0)+1\right)+\mathfrak{P}(y_0)\right)\right),\\\nonumber
\mathcal{A}_3&=&8 \mathfrak{P}(y_0)
\left(\mathcal{F}_-'(y_0)-\mathcal{F}_+'(y_0)\right)+\mathfrak{S}
(y_0) \left(y_0 \mathcal{F}_-''(y_0)+4 \mathcal{F}_-'(y_0)-y_0
\mathcal{F}_+''(y_0)-4 \mathcal{F}_+'(y_0)\right),\\\nonumber
\mathcal{A}_4&=&8 \mathfrak{P}(y_0)
\left(\mathcal{F}_-'(y_0)-\mathcal{F}_+'(y_0)\right)+\mathfrak{S}
(y_0) \left(y_0 \mathcal{F}_-''(y_0)+4 \mathcal{F}_-'(y_0)-y_0
\mathcal{F}_+''(y_0)-4 \mathcal{F}_+'(y_0)\right),\\\nonumber
\mathcal{A}_5&=&4 \mathfrak{S} (y_0) (\mathfrak{P}(y_0)+\mathfrak{S}
(y_0)) \mathfrak{P}'(y_0)-10 \mathfrak{P}(y_0) \mathfrak{S} (y_0)-12
\mathfrak{P}(y_0)^2-\mathfrak{S} (y_0)^2.
\end{eqnarray}

\begin{figure}\centering
\epsfig{file=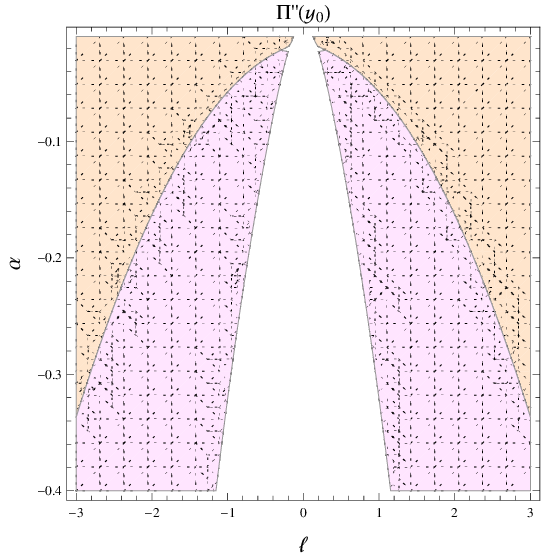,width=.33\linewidth}\epsfig{file=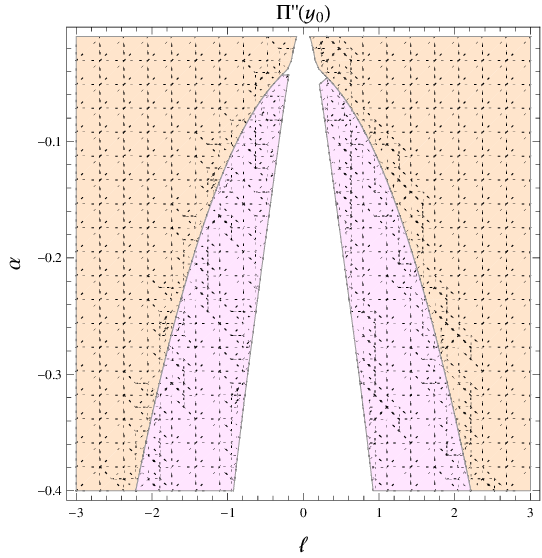,width=.33\linewidth}\epsfig{file=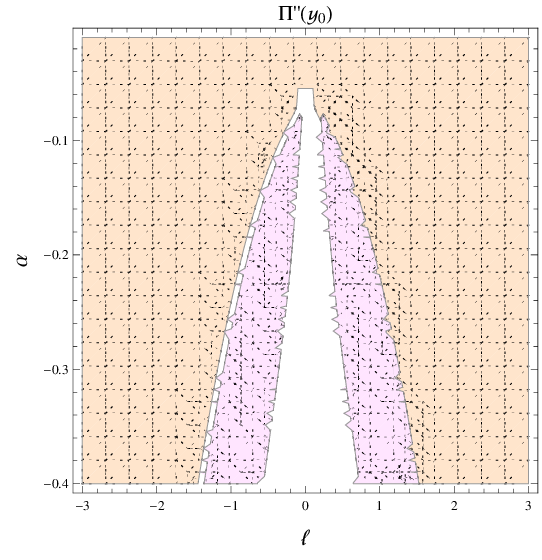,width=.33\linewidth}
\caption{\label{p1} Plots of $\Pi''(y_{0})$ along $\alpha$ and
$\ell$ for dark energy for various choices of $c_{1}$ as
$c_{1}=0.3$ (first plot), $c_{1}=0.5$ (second plot) and $c_{1}=0.9$
(third plot) for $\omega=-1,y_0=\left(l^m+1\right)^{1/m},h_{0}=1,K_{2} =0.1,c_{2} =0.5,m=2$.}
\epsfig{file=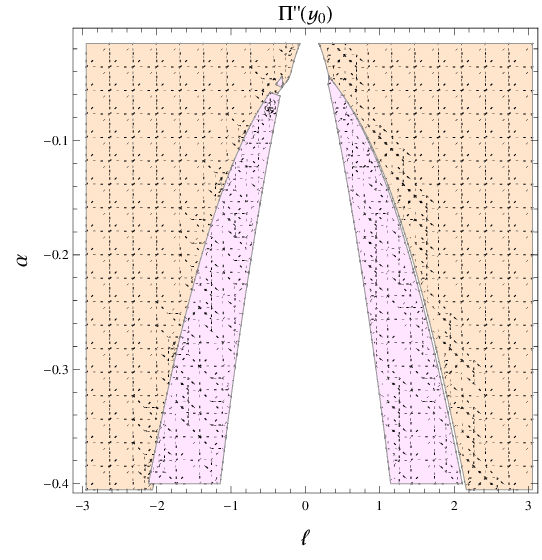,width=.33\linewidth}\epsfig{file=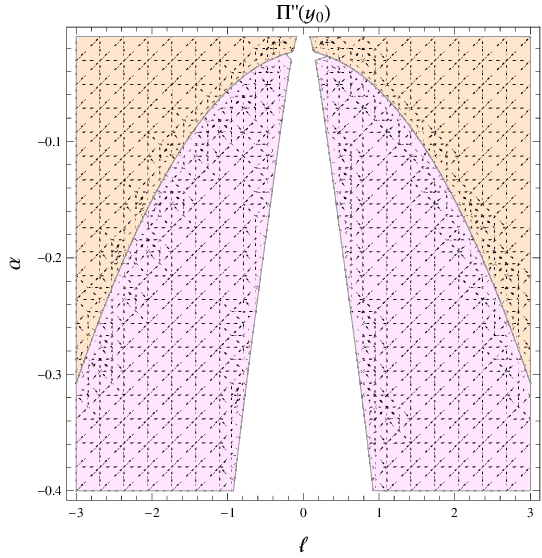,width=.33\linewidth}\epsfig{file=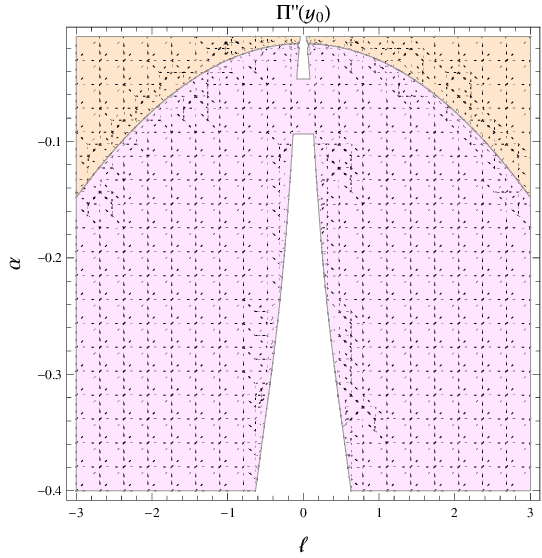,width=.33\linewidth}
\caption{\label{p2} Plots of $\Pi''(y_{0})$ along $\alpha$ and
$\ell$ for quintessence energy for various choices of $c_{1}$ as
$c_{1}=0.3$ (first plot), $c_{1}=0.5$ (second plot) and $c_{1}=0.9$
(third plot) for $\omega=-2/3,y_0=\left(l^m+1\right)^{1/m},h_{0}=1,K_{2}
=0.1,c_{2} =0.5,m=2$.}
\epsfig{file=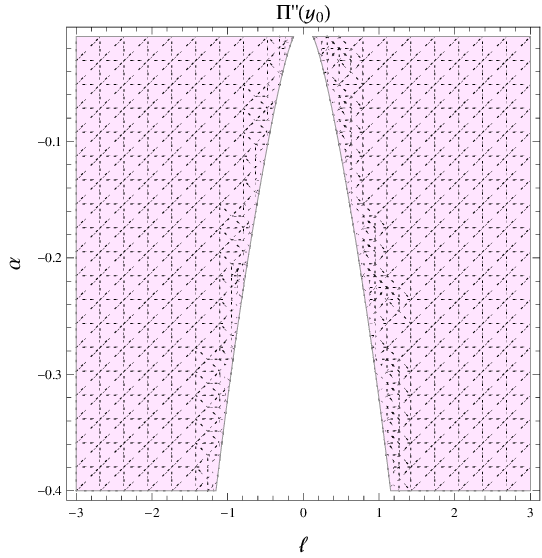,width=.33\linewidth}\epsfig{file=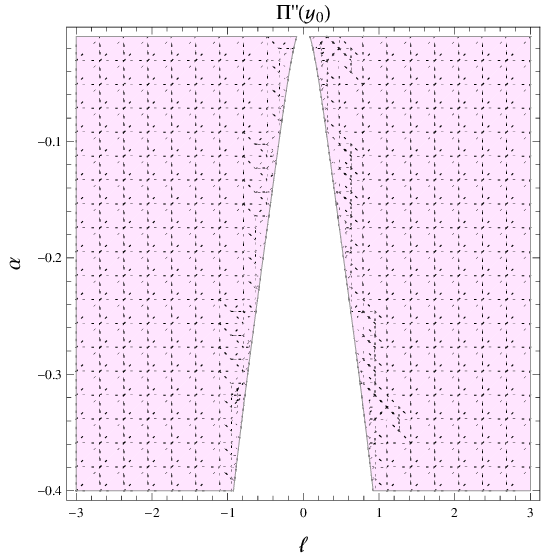,width=.33\linewidth}\epsfig{file=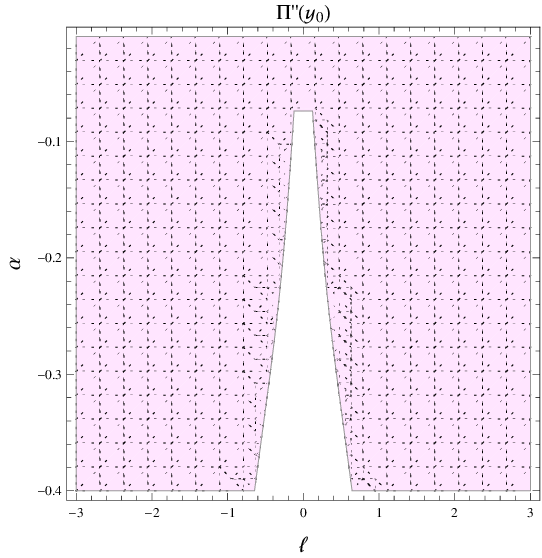,width=.33\linewidth}
\caption{\label{p3} Plots of $\Pi''(y_{0})$ along $\alpha$ and
$\ell$ for phantom energy for various choices of $c_{1}$ as
$c_{1}=0.3$ (first plot), $c_{1}=0.5$ (second plot) and $c_{1}=0.9$
(third plot) for $\omega=-2,y_0=\left(l^m+1\right)^{1/m},h_{0}=1,K_{2} =0.1,c_{2} =0.5,m=2$.}
\end{figure}
\begin{figure}\centering
\epsfig{file=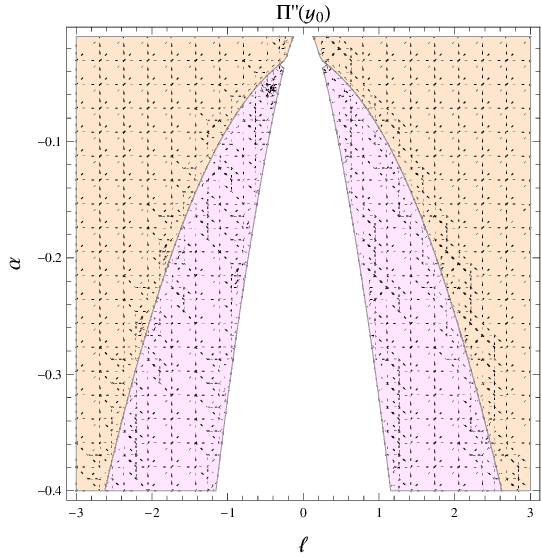,width=.33\linewidth}\epsfig{file=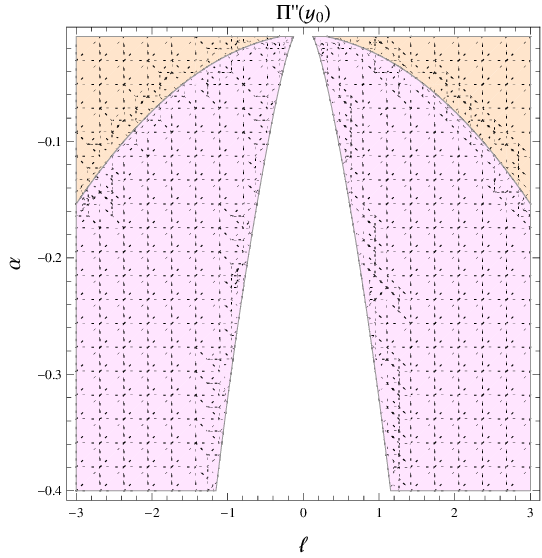,width=.33\linewidth}\epsfig{file=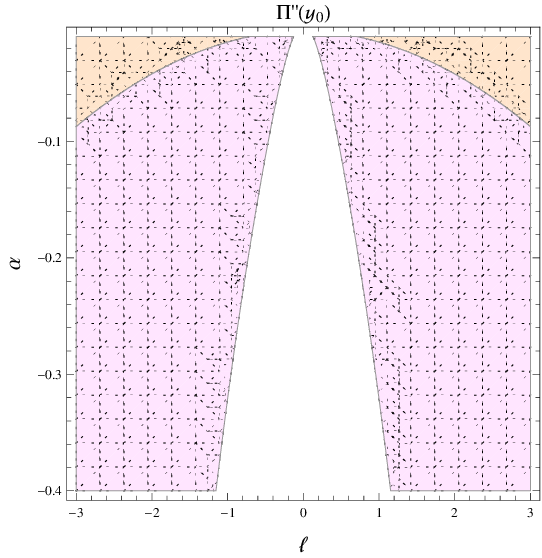,width=.33\linewidth}
\caption{\label{p4} Plots of $\Pi''(y_{0})$ for dark energy for
$h_{0}=0.8$ (first plot), $h_{0}=1.5$ (second
plot) and $h_{0}=2$ (third plot) for
$\omega=-1,y_0=\left(l^m+1\right)^{1/m},c_{1}=0.3,K_{2} =0.1,c_{2}
=0.5,m=2$.}
\epsfig{file=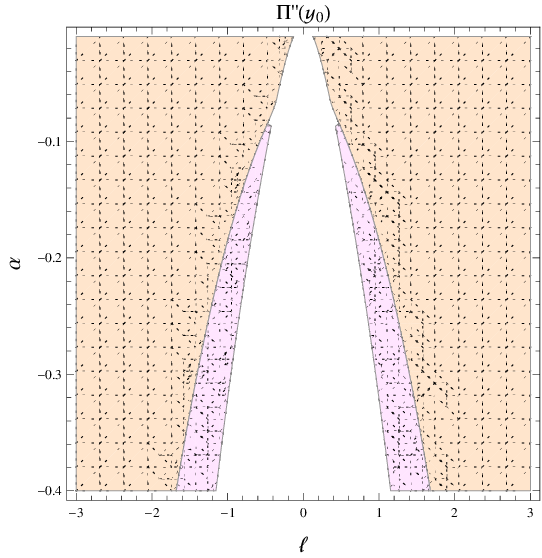,width=.33\linewidth}\epsfig{file=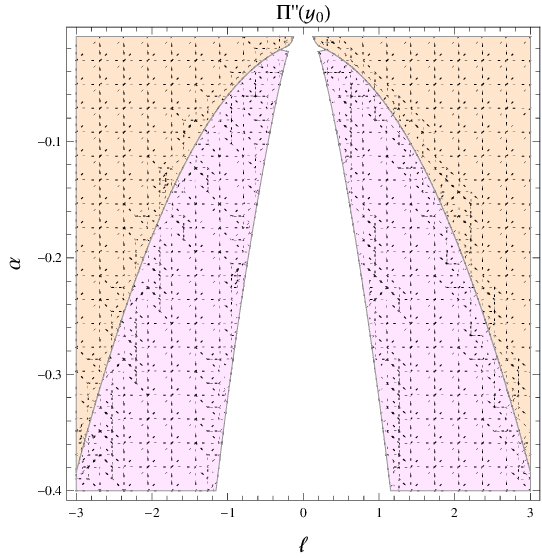,width=.33\linewidth}\epsfig{file=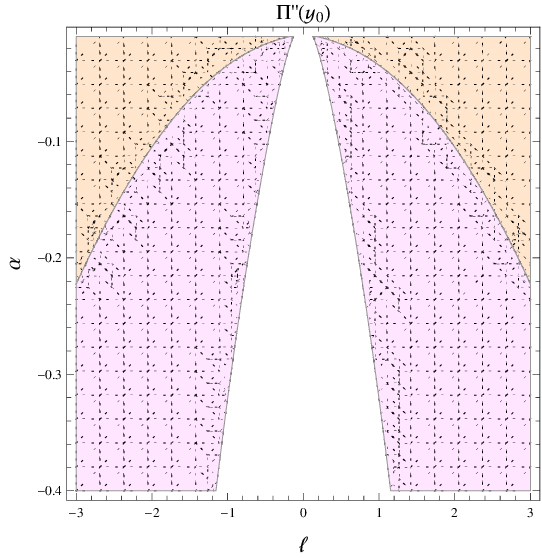,width=.33\linewidth}
\caption{\label{p5} Plots of $\Pi''(y_{0})$ for quintessence energy
for $h_{0}=0.8$ (first plot), $h_{0}=1.5$
(second plot) and $h_{0}=2$ (third plot) for
$\omega=-2/3,y_0=\left(l^m+1\right)^{1/m},c_{1}=0.3,K_{2}
=0.1,c_{2} =0.5,m=2$.}
\epsfig{file=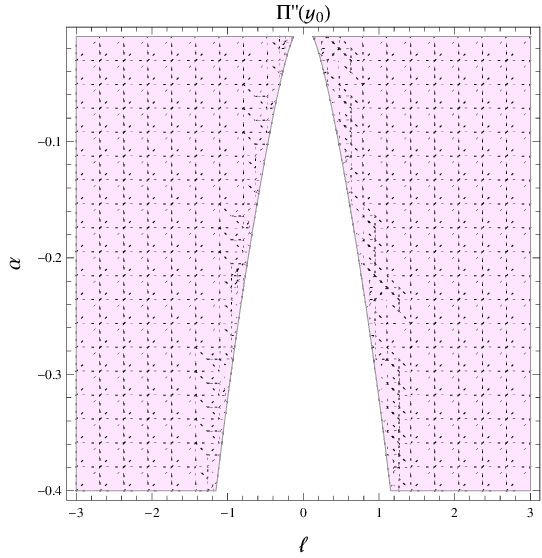,width=.33\linewidth}\epsfig{file=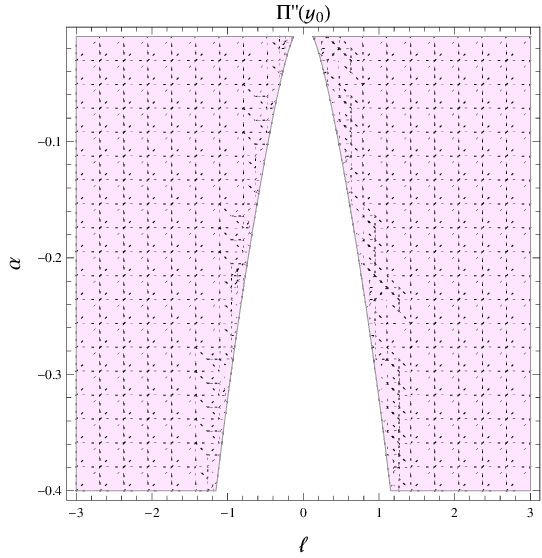,width=.33\linewidth}\epsfig{file=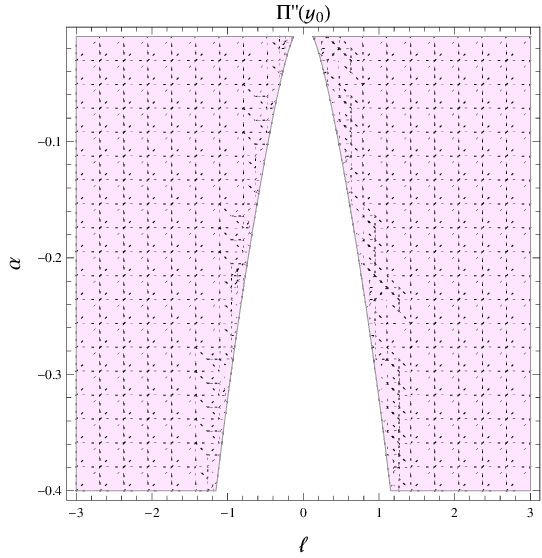,width=.33\linewidth}
\caption{\label{p6} Plots of $\Pi''(y_{0})$ for phantom energy for $h_{0}=0.8$ (first plot), $h_{0}=1.5$
(second plot) and $h_{0}=2$ (third plot) for
$\omega=-2,y_0=\left(l^m+1\right)^{1/m},c_{1}=0.3,K_{2} =0.1,c_{2}
=0.5,m=2$.}
\end{figure}

\begin{figure}\centering
\epsfig{file=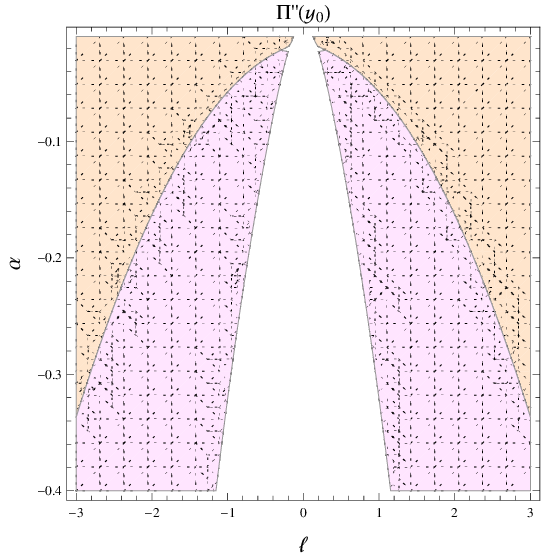,width=.33\linewidth}\epsfig{file=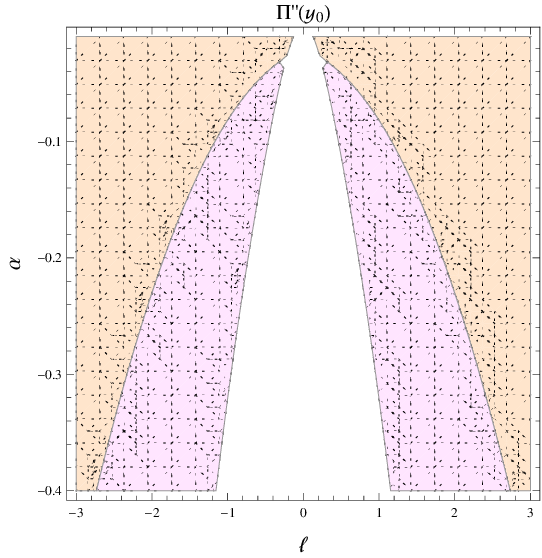,width=.33\linewidth}\epsfig{file=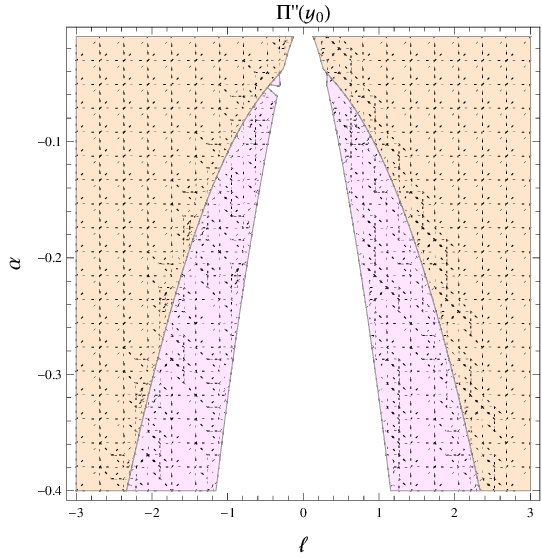,width=.33\linewidth}
\caption{\label{p10} Plots of $\Pi''(y_{0})$ for dark energy for
$K_{2} =0.1$ (first plot), $K_{2} =0.12$ (second plot) and $K_{2}
=0.14$ (third plot) for
$\omega=-1,y_0=\left(l^m+1\right)^{1/m},c_{1}=0.3,h_{0}=1,c_{2}
=0.5,m=2$.}
\epsfig{file=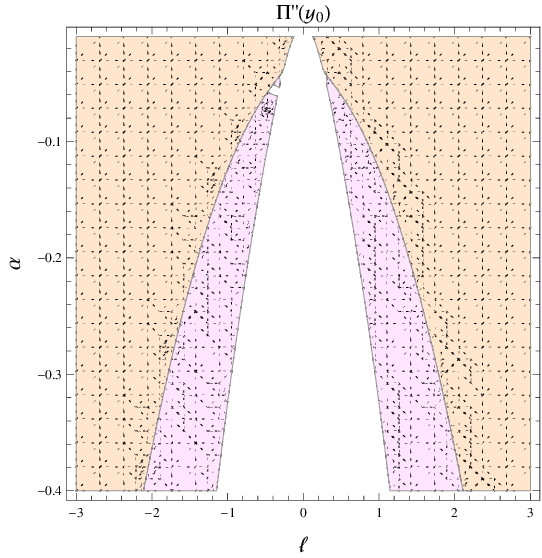,width=.33\linewidth}\epsfig{file=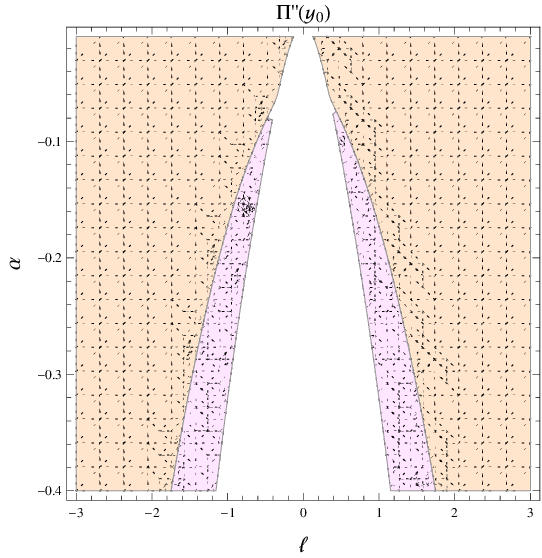,width=.33\linewidth}\epsfig{file=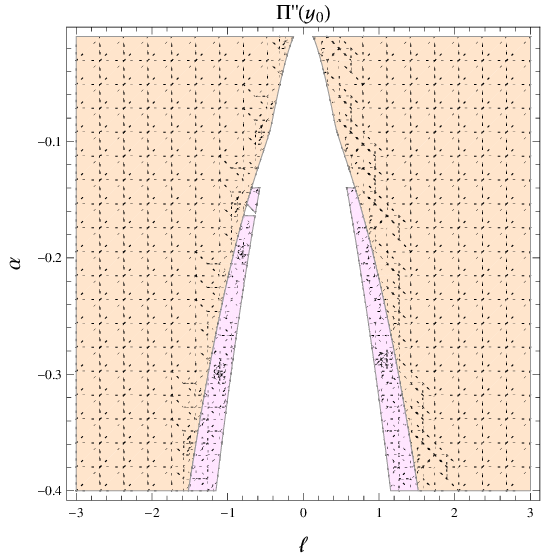,width=.33\linewidth}
\caption{\label{p11} Plots of $\Pi''(y_{0})$ for quintessence energy
for $K_{2} =0.1$ (first plot), $K_{2} =0.12$ (second plot) and
$K_{2} =0.14$ (third plot) for
$\omega=-2/3,y_0=\left(l^m+1\right)^{1/m},c_{1}=0.3,h_{0}=1,c_{2}
=0.5,m=2$.}
\epsfig{file=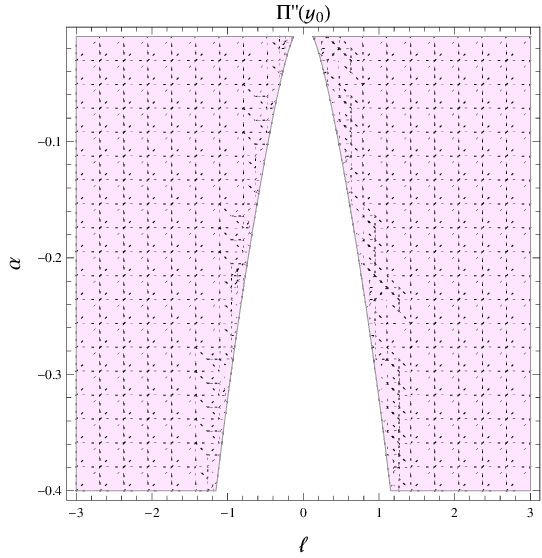,width=.33\linewidth}\epsfig{file=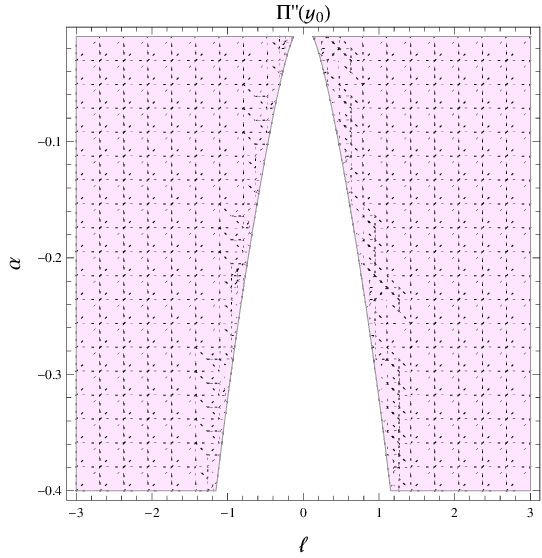,width=.33\linewidth}\epsfig{file=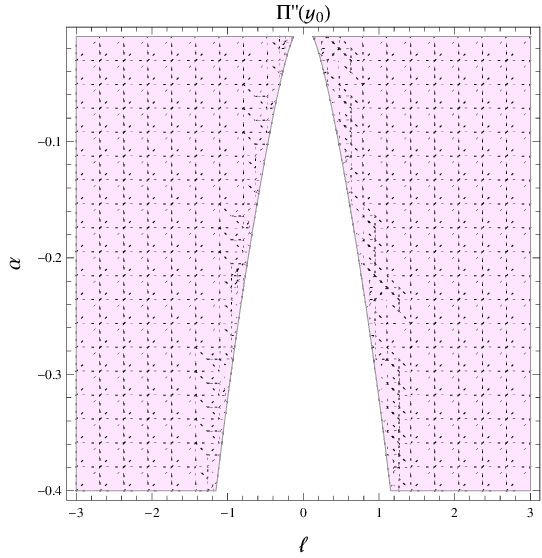,width=.33\linewidth}
\caption{\label{p12} Plots of $\Pi''(y_{0})$ for phantom energy for
$K_{2} =0.1$ (first plot), $K_{2} =0.12$ (second plot) and $K_{2}
=0.14$ (third plot) for
$\omega=-2,y_0=\left(l^m+1\right)^{1/m},c_{1}=0.3,h_{0}=1,c_{2}
=0.5,m=2$.}
\end{figure}

\subsection{Phantomlike EoS}

The stable thin-shell configuration is a highly significant subject in the realms of astrophysics and cosmology, since it assists in the inspection of valid gravastar results. EoS is essential for examining the impacts of distinct kinds of matter constituents at the hypersurface on the stable state of a thin shell. Among various speculations that can narrate the exotic substance, one is a phantom-like EoS, that is portrayed as
\begin{equation}\label{17kk}
\mathfrak{P}=\omega \mathfrak{S}(y),
\end{equation}
where the parameter of EoS $\omega<0$. The various
bounds of EoS parameter characterize the distinct kinds of
matter constituents. It exhibits dark energy, quintessence, and
phantom energy state if $\omega<-1/3$, $0>\omega>-1/3$ and
$\omega<-1$, respectively. To explore the thin-shell stability for three distinct forms of EoS
parameter, i.e., dark energy, quintessence, and phantom energy like
matter constituents, we observe the graphical behavior of the second
derivative of effective potential as presented in Eq. (\ref{sad}). For
simplification, we can consider
$y=\left(S_{0}^{m}+\ell^{m}\right)^{1 / m}$, in which $\ell$ depicts the length
factor and $m$ acts as a constant. We scrutinize the outcome of distinct
factors on the stable state of the shell for
different types of energy contents. Some detailed outcomes are given
below:

\begin{itemize}
\item Figs. (\ref{p1})-(\ref{p3}) are used to explore the impact of $c_{1}$ on the stability of gravastars for dark, quintessence and phantom energies, respectively. Here, we use the regional plots to portray the stable and unstable states of the shell. The stable regions are denoted with the pink doted region and the unstable region is represented with the grey doted region. It is noted that the stability is reduced by increasing the parameter $c_{1}$ for the dark energy distributions see Fig. (\ref{p1}). In the presence of quintessence energy, this parameter behaves differently. For such matter contents, the stability regions become greater for higher values of  $c_{1}$ see Fig. (\ref{p2}). For the phantom energy, the created geometry displays maximal stable nature for smaller as well as higher values of $c_{1}$ as compared to the dark as well as quintessence energy contents see Fig. (\ref{p3}).

\item Figs. (\ref{p4})-(\ref{p6}) are used to observe the stable/unstable phase of the obtained structure for different choices of $h_{0}$ having distinct kinds of matter contents. For the selection of dark and quintessence matter distribution, the developed structure expresses maximum stable behavior for greater choices of $h_{0}$ see Fig. (\ref{p4}) and (\ref{p5}). For the phantom energy, the gained structure only exhibits stable conduct and the stable regions increase by increasing $h_{0}$.

\item Figs.(\ref{p10})-(\ref{p12}) are devoted to explain the impact of $K_{2}$. It is noted that the smaller values of $K_{2}$ are only suitable for the stable behavior of the gravastar structure see Fig.(\ref{p10}). For phantom energy system shows more unstable behavior as compared to the dark energy see Fig.(\ref{p11}). For the phantom energy again system is expresses maximum stable behavior for every choice of the $K_{2}$ see Fig.(\ref{p12}).
\end{itemize}

\section{Some Physical Attributes}

\begin{figure}\centering
\epsfig{file=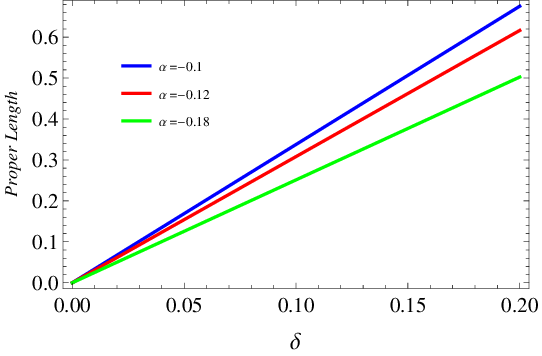,width=.5\linewidth}\epsfig{file=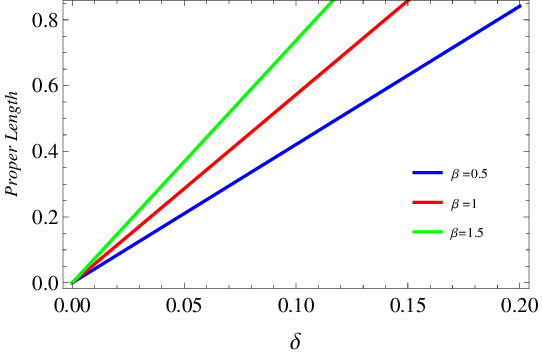,width=.5\linewidth}
\caption{\label{g1} Proper length along $\delta$ for various choices of $\alpha$ (left plot) and $c_{1}$ (right plot)  with
$h_{1}=1,\alpha =0.5,Y =0.5,c_{2} =1,z=10;$.}
\end{figure}

The physical properties of charged gravastars like appropriate length, energy, and entropy with CKVs, are the main topics of this section. The shell's proper length, denoted by $\delta$, and thickness are discussed here. The thickness of the shell is a tiny real positive value, $\delta$, such that $0<\delta\ll1$. The parameters $y$ and $y+\delta$ specify the shell's bounds. The appropriate length of the shell is displayed mathematically as follows \cite{qm}
\begin{equation}\label{19}
l=\int_{y}^{y+\delta}\sqrt{e^{Y(r)}}=\int_{y}^{y+\delta}
\sqrt{\frac{K_{2} r^{\frac{4 (c_{1} +1)}{c_{1} +4}} ((3 c_{1}
+2) r)^{\frac{8}{3 c_{1} +2}} \left(\alpha  \left(6 r^2-3 (c_{1}
+1)\right)+2 (c_{1} +1) \left(r^2-1\right) r^2 c_{2} \right)}{12
\alpha  (c_{1} +1) \left(r^{\frac{4 (c_{1} +1)}{c_{1} +4}}-r^2 ((3
c_{1} +2) r)^{\frac{8}{3 c_{1} +2}}\right)}}dr.
\end{equation}
To resolve the aforementioned intricate integration, we presume that the
\begin{equation}\label{19f}\sqrt{\frac{K_{2} r^{\frac{4
(c_{1} +1)}{c_{1} +4}} ((3 c_{1} +2) r)^{\frac{8}{3 c_{1} +2}}
\left(\alpha  \left(6 r^2-3 (c_{1} +1)\right)+2 (c_{1} +1)
\left(r^2-1\right) r^2 c_{2} \right)}{12 \alpha  (c_{1} +1)
\left(r^{\frac{4 (c_{1} +1)}{c_{1} +4}}-r^2 ((3 c_{1} +2)
r)^{\frac{8}{3 c_{1} +2}}\right)}}=\frac{dB(r)}{dr},
\end{equation}
as
\begin{eqnarray}\nonumber
l&=&\int_{y}^{y+\delta}\frac{dB(r)}{dr}dr=B(y+\delta)-B(y)\approx
\delta
\frac{dB(r)}{dr}|_{r=y}\\
&=&\delta\sqrt{\frac{K_{2}
y^{\frac{4 (c_{1} +1)}{c_{1} +4}} ((3 c_{1} +2) y)^{\frac{8}{3 c_{1}
+2}} \left(\alpha  \left(6 y^2-3 (c_{1} +1)\right)+2 (c_{1} +1)
\left(y^2-1\right) y^2 c_{2} \right)}{12 \alpha  (c_{1} +1)
\left(y^{\frac{4 (c_{1} +1)}{c_{1} +4}}-y^2 ((3 c_{1} +2)
y)^{\frac{8}{3 c_{1} +2}}\right)}}.
\end{eqnarray}
$\delta$ is a small real positive constant whose square and higher powers can be neglected. The shell's thickness and the proper length can therefore be correlated and examined. Both the shell radius and the physical parameters influence this relationship. It is found that the proper length of the shell increases by increasing the shell's thickness see Fig. (\ref{g1}). Also, the physical parameter $\alpha$ greatly affects the proper length. It is noted that the proper length becomes smaller by enhancing $\alpha$ negatively. Also, this graphical analysis explains the direct relation between the proper length and thickness of the shell. It is also observed that the parameter affects the proper length. It is found that the proper becomes greater by enhancing the parameter $c_{1}$ see the right plot of Fig. (\ref{g1}).

\begin{figure}\centering
\epsfig{file=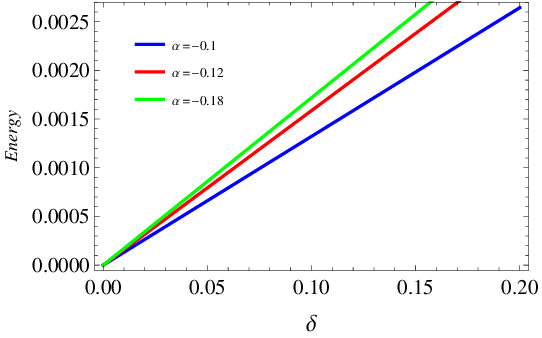,width=.5\linewidth}\epsfig{file=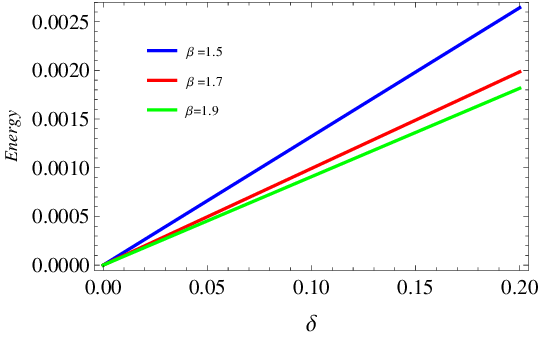,width=.5\linewidth}
\caption{\label{g2} Energy contents
 along $\delta$ for various choices of $\alpha$ (left plot) and $c_{1}$ (right plot)  with
$h_{1}=1,\alpha =0.5,Y =0.5,c_{2} =1,z=10;$.}
\end{figure}

\begin{figure}\centering
\epsfig{file=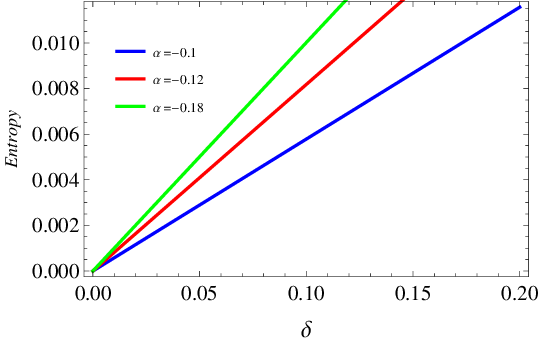,width=.5\linewidth}\epsfig{file=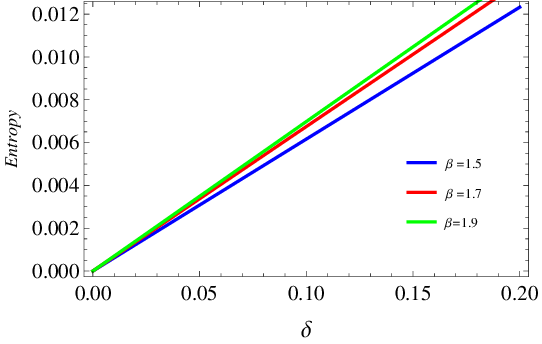,width=.5\linewidth}
\caption{\label{g3} Entropy of shell  along $\delta$ for distinct choices of $\alpha$ (left plot) and $c_{1}$ (right plot)  with
$h_{1}=1,\alpha =0.5,Y =0.5,c_{2} =1,z=10;$.}
\end{figure}

A non-attractive force and a negative energy zone exist in the internal region of a gravastar, because the substance follows the equation of state $p=-\rho$. Finding the appropriate length as
\cite{qm}
\begin{equation}\label{25}
\textbf{E}=\int_{y}^{y+\delta}4\pi r^2 \rho(r)dr=\delta \frac{\alpha
y^{-\frac{4 (c_{1} +1)}{c_{1} +4}} \left(6 \left(c_{1} ^2+3 c_{1}
+2\right) h_{1}+(c_{1} +4) K_{2} y^{\frac{4 (c_{1}
+1)}{c_{1} +4}}\right)}{4 (c_{1} +1) (c_{1} +2) (c_{1} +4) K_{2}
^2}.
\end{equation}
The final shell energy expression is dependent on the physical parameters of gravity and the thickness and radius of the shell.

It is found that the energy of the shell reduces by enhancing the shell's thickness see Fig. (\ref{g2}). Also, the physical parameter $\alpha$ greatly affects the energy. It is noted that the energy becomes larger as  $\alpha$ increases negatively. Also, this graphical analysis explains the inverse relation between the energy of the shell and the $c_{1}$. For higher values of $c_{1}$, the energy of the shell becomes smaller see the right plot of Fig. (\ref{g2}).

The degree of disturbance or disruption in a structure of geometry is able to be determined using the amount of entropy of the structure. To measure the randomness of gravastar, the shell's entropy is examined. Mazur and Mottola's concept is utilized to get the entropy equation of a gravastar thin-shell as
 \cite{qm}
\begin{equation}\label{22a}
S=\int_{y}^{y+\delta}4\pi r^2 j(r) \sqrt{e^{Y(r)}}dr.
\end{equation}
The entropy density at a given local temperature is computed as
\begin{equation}\label{23a}
j(r)=\frac{\eta K_B}{\hbar}\sqrt{\frac{p(r)}{2\pi}},
\end{equation}
with $\eta$ as a dimensionless factor. Here, we consider
Planck units $(K_B = 1 = \hbar)$ yielding the following form of the shell's entropy
\cite{qm}
\begin{equation}\label{24a}
S= \delta \frac{y^2 \sqrt{\frac{y^{-\frac{6 (c_{1} +2)}{c_{1} +4}}
\left(6 \alpha  (c_{1} +1) (c_{1} +2) h_{1}+\alpha  (c_{1} +4)
K_{2} y^{\frac{4 (c_{1} +1)}{c_{1} +4}}\right)}{(c_{1} +1)
(c_{1} +2) (c_{1} +4) K_{2}}} \sqrt{\frac{K_{2} y^{\frac{4
(c_{1} +1)}{c_{1} +4}} ((3 c_{1} +2) y)^{\frac{8}{3 c_{1} +2}}
\left(\alpha  \left(6 y^2-3 (c_{1} +1)\right)+2 (c_{1} +1)
\left(y^2-1\right) y^2 c_{2} \right)}{\alpha  (c_{1} +1)
\left(y^{\frac{4 (c_{1} +1)}{c_{1} +4}}-y^2 ((3 c_{1} +2)
y)^{\frac{8}{3 c_{1} +2}}\right)}}}{2 \sqrt{6}}.
\end{equation}
Moreover, the shell's entropy should be proportional to $\delta$.
It is gained that the energy of the shell reduces by enhancing the shell's thickness see Fig. (\ref{g3}). It is noted that the entropy of the shell increases by enhancing $\alpha$ negatively. For higher values of $c_{1}$, the entropy of the shell becomes larger.

\section{Conclusion}\label{sec:6}

In this paper, we scrutinize the gravastar geometry in $f(\mathcal{T},T)$ gravity, an extended teleparrlel gravity with spherically symmetric space-time and an isotropic fluid source. By using the conformal symmetry for the spherically symmetric space-time, we want to investigate gravastar. As far as we are aware, this is the initial effort to look into the gravastar $f(\mathcal{T},T)$ theory. To sum up, the investigation of gravastar geometry has uncovered discrete areas with disparate equations of state ($p=\omega\rho$, where $\omega=-1,1,0$) and conformal coordinates of death. The Kretschmann scalar confirmed regularity and the determination of precise solutions for the interior, intermediate, and outer areas. Gravastars' overall geometry was formed by joining these regions with junction conditions, taking into account two possible external geometry possibilities. The stability and physical properties of charged gravastars were investigated; stability restrictions in terms of the coupling constant and conformal parameters were discussed. The assessment of the produced structure's appropriate length, energy content, and entropy adds to our knowledge about gravastars and their characteristics. The detail discussion are given below:
\begin{itemize}
\item The stability of gravastars is closely connected to the parameters $c_{1}$, $h_{0}$, and $K_{2}$, which represent the specific sort of energy content within the system see Figs. (\ref{p1})-(\ref{p12}). The behavior of gravastar structures exhibited substantial variations depending on the dominant presence of either dark, quintessence, or phantom energy. Increasing the value of $c_{1}$ generally decreases the stability of dark energy, but it improves the stability of quintessence energy and results in the highest level of stability for phantom energy. Similarly, larger values of $h_{0}$ enhanced stability in structures containing dark and quintessence energy, whereas structures with phantom energy remain stable regardless of the chosen value of $h_{0}$. Furthermore, it has been observed that gravastar structures tend to be more stable when the values of $K_{2}$ are lower. In particular, systems with phantom energy show higher levels of instability compared to systems with dark energy. These findings emphasize the intricate interaction among many energy components and characteristics in affecting the stability of gravastars.
\item The association between the shell thickness and its suitable length has been analyzed in detail; physical characteristics and the shell radius are important factors in defining this correlation. As shown in Fig. (\ref{g1}), it is seen that the shell's proper length increases with the shell's thickness. The proper length is strongly influenced by the physical parameter $\alpha$; a negative rise in $\alpha$ results in a drop in the proper length. Furthermore, as can be observed in the right plot of Fig. (\ref{g1}), the parameter $c_{1}$ also influences the proper length, with larger values resulting in a longer proper length. Moreover, as the thickness increases, the shell's energy drops, as shown in Fig. (\ref{g2}), where $\alpha$ and $c_{1}$ are critical factors in determining the energy levels. Based on Fig. (\ref{g3}), it is determined that the shell's entropy is proportional to $\delta$, increasing with a negative enhancement in $\alpha$ and for greater values of $c_{1}$.
\end{itemize}
These results offer important new understandings of how shell parameters and physical attributes interact to define the shell's properties within gravastar structures.

\section*{Acknowledgement}

F. Javed greets the financial assistance received for his postdoctoral fellowship under Grant No. YS304023917 at Zhejiang Normal University.

\end{document}